\documentclass[fleqn,twoside]{article}
\usepackage{espcrc2}
\usepackage{graphicx}

\newcommand{\AmS}{{\protect\the\textfont2
  A\kern-.1667em\lower.5ex\hbox{M}\kern-.125emS}}
\newcommand{\m}{\rm \,m}
\newcommand{\mm}{\rm \,mm}

\newcommand{\sr}{\rm \,sr}

\newcommand{\GeV}{\rm \,GeV}

\newcommand{\numu}{$\nu_{\mu}$ }

\newcommand{\nutau}{$\nu_{\tau}$ }

\newcommand{\lsim}{\lower .5ex\hbox{$\buildrel < \over {\sim}$}}
\newcommand{\gsim}{\lower .5ex\hbox{$\buildrel > \over {\sim}$}}

\title{MACRO results on atmospheric neutrinos}
\author{G. Giacomelli\address[BO]{Dipartimento di Fisica and INFN,\\ 
        viale C. Berti-Pichat, 6/2, I-40127 Bologna, Italy\\
        \textit{Paper presented at the NOW 2004 Workshop, Conca Specchiulla, Otranto, Italy, September 2004.}      }  and
        A. Margiotta\addressmark[BO]\\
        for the MACRO Collaboration\thanks{see Ref. [1] for a list of MACRO Authors
    and  Institutions }} 
\begin{document}
\begin{abstract}
We discuss the final results of the MACRO experiment on atmospheric neutrino oscillations. The data concern event topologies with average neutrino energies of $\sim 3$ and $\sim 50$ GeV. Multiple Coulomb Scattering of the high energy muons was used to estimate the neutrino energy event by event. The angular distributions, the $L/E_\nu$ distribution, the particle ratios and the absolute fluxes  all favor $\nu_\mu \rightarrow \nu_\tau$ oscillations with maximal mixing and $\Delta m^2 \simeq 0.0023 \: \rm eV^2$. Emphasis is given to measured ratios which are not affected by Monte Carlo (MC) absolute normalization; a discussion  is made on MC uncertainties. A preliminary search for possible Lorentz invariance violation contributions to atmospheric neutrino oscillations is presented and discussed.
\vspace{1pc}
\end{abstract}
\maketitle
\section{Introduction}
\label{intro}
MACRO was a large area multipurpose underground detector \cite{r1} designed to search for rare events and rare phenomena in the penetrating cosmic radiation \cite{r2}.  It was located in Hall B of  the  Gran Sasso Lab at
an average rock  overburden of 3700 m.w.e.; it started data taking with part of the apparatus in \( 1989 \); it was
completed in \( 1995 \) and   run in its final configuration until the end of 2000.  The detector had dimensions of \( 76.6\times12 \times9 .3{\m }^{3} \) and provided a total acceptance to an isotropic flux of particles of \(\sim 10,000{\m }^{2}{\sr } \); vertically it was divided into a lower part, which contained 10 horizontal layers of streamer tubes, 7 of rock absorbers and 2 layers of liquid  scintillators, and an upper part which contained the
electronics and was  covered by 1 layer of scintillators and 4 layers of streamer tubes. The sides were covered with 1 vertical layer of scintillators and 6 of limited streamer tubes.

MACRO detected upgoing $\nu_\mu$'s via charged current interactions, \(\nu _{\mu } \rightarrow \mu \); upgoing muons were identified with the streamer tube system (for tracking) and the scintillator system (for time-of-flight). The events measured and expected for the 3 measured topologies, deviate from Monte Carlo (MC) expectations without oscillations; these deviations and the $L/E_\nu$ distribution point to the same  ${\nu_\mu \rightarrow  \nu_\tau}$  oscillation scenario \cite{r2}-\cite{r8}. Here we also present the results of a preliminary search for possible Lorentz invariance violation contributions to atmospheric neutrino oscillations.  
\section{Atmospheric neutrinos. Monte Carlos}
\begin{figure}[h]
\begin{center}
\resizebox{0.7\hsize}{!}{%
  \includegraphics{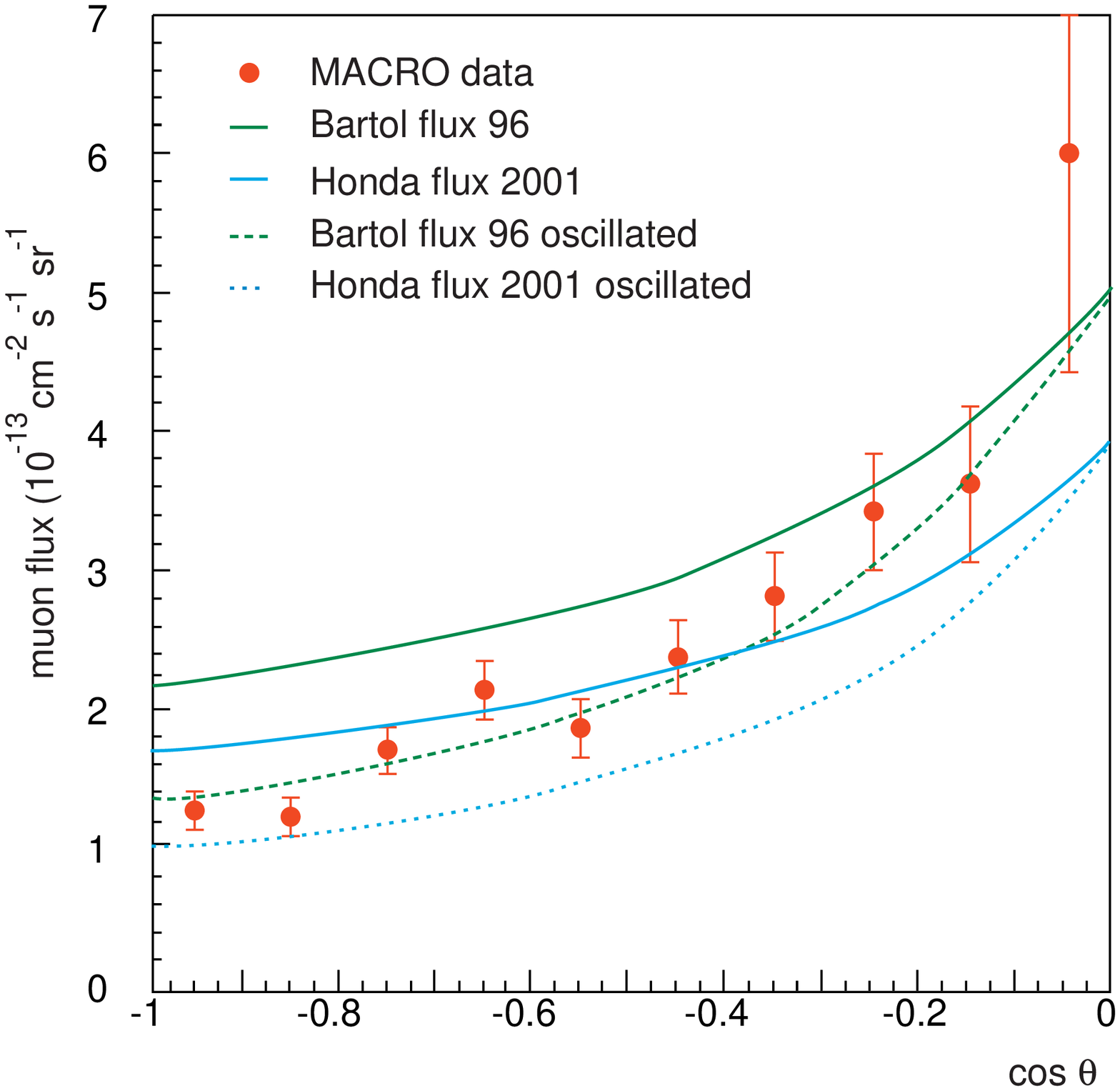} }
  \resizebox{0.4\vsize}{!}{%
  \includegraphics{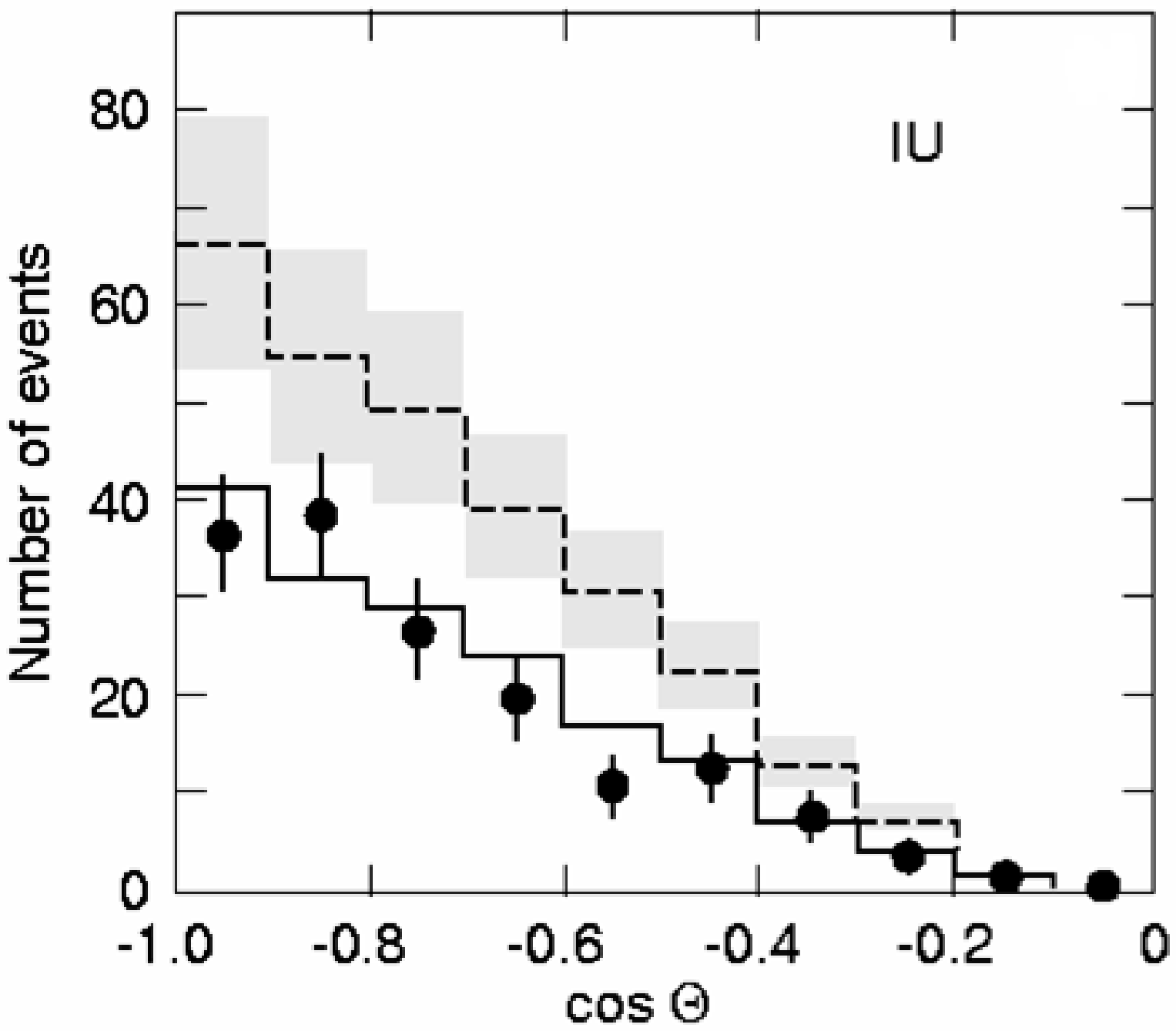}
}
\end{center}
\vspace{-1.2cm}
 \caption{\label{fig2}\small Zenith distributions for MACRO data (black
  points) for (a) upthroughgoing (top) and (b) semicontained muons.
In (a) there are comparisons between the predictions of old and new MCs with and without oscillations. In (b) the dashed line is the no-oscillation Bartol96 MC (with an error band); the solid line is for 
 \( \nu _{\mu }\rightarrow  \nu _{\tau } \) oscillations  with $\sin^{2} 2\vartheta = 1$ and $\Delta  m^{2} = 2.3 \cdot 10^{-3}$ eV$^{2}$. }
\end{figure}

The measured upthroughgoing muon data of Fig. \ref{fig2}a were compared with different MC simulations. In the past we used the $\nu$ flux computed by the Bartol96 group \cite {bartol}. The systematic uncertainty in the predicted flux was estimated at $\pm 17 \:\%$;  this is mainly a scale error that does not change the shape of the angular distribution. A similar MC (Honda95)\cite{honda95}  was used by the SuperK Collaboration (SK) \cite{sk98}.
Recently new improved MC predictions for neutrino fluxes were made available by the Honda (HKKM01) \cite{honda01} and FLUKA \cite {fluka} groups. They include three dimensional calculations of hadron production and decay and of neutrino  interactions, improved hadronic model and new fits of the primary cosmic ray flux. The two MC yield predictions for
the non oscillated and oscillated \numu fluxes equal to within few \% \cite{r8}. The shapes of the angular distributions for oscillated and non oscillated Bartol96, new FLUKA and new Honda fluxes are the same to within few \%.
The absolute values of our data are higher than  those  predicted by the new FLUKA and Honda MC, Fig. \ref{fig2}. A similar situation is found in the  new SK data \cite{hayato}. 
The L3C and BESS cosmic ray results presented at this workshop \cite{r13} lead to $\nu$ fluxes in better agreement with Bartol96 and Honda95 predictions. The evidence for neutrino oscillations is  mainly due to the shape  of the angular distribution and this is the same in all MCs. Also the ratios of the medium to high energy measurements and of the two low energy data  samples  are MC independent. Our data suggest that the FLUKA normalization should be raised by  $25 \%$ at $E_\nu \sim 50 \; GeV$ and by $12 \%$ at $E_\nu \sim 3 \; GeV$.
\begin{figure}[th]
\begin{center}
\resizebox{0.9\hsize}{!}{%
  \includegraphics{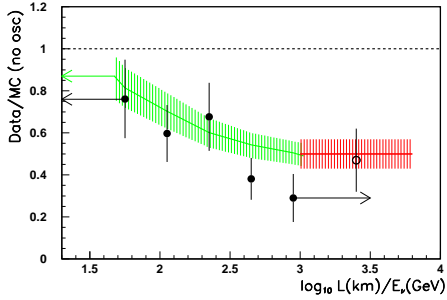}
  }
\end{center}
\vspace{-1.2cm}
\caption{\label{fig3}\small  Ratio (Data/MC Bartol96) versus the estimated $L/E_{\nu}$ for the upthroughgoing
muons  (black circles) and the semicontained up-$\mu$ (open circle). The  horizontal dashed line at
Data/MC=1 is the  expectation for no oscillations.}
\end{figure}
\section{MACRO results on atmospheric $\nu$'s}
The {\it upthroughgoing muons } come from $ \nu _{\mu } $ interactions in the rock below the detector; muons with \(E_{\mu }>1\GeV  \) cross the whole detector. The corresponding \( \nu _{\mu } \)'s have a median energy of 50 GeV.  Many  systematic effects and backgrounds  were studied \cite{r3,r8}. The data, Fig. \ref{fig2}a, agree in shape and  absolute value with  the oscillated Bartol96 MC, for $\Delta m^2 $ = 0.0023 eV$^2$.\\
\textit{\numu $\rightarrow$ \nutau versus \numu $\rightarrow \nu_s$}. 
The ratio $R_1 = $ Vertical/Horizontal $= N(-1<cos\theta < -0.7) / N(-0.4 < cos\theta < 0)$ was  used to test the \numu $\rightarrow \nu_s$ oscillation hypothesis versus \numu $\rightarrow$ \nutau \cite{r2} \cite{r6} \cite{r8}. 
The  \numu $\rightarrow \nu_s$ oscillations (with any mixing) are excluded at  $99.8\%$ c.l. with respect to \numu
$\rightarrow$ \nutau oscillations with  maximal mixing \cite{r8}.\\
\textit{Oscillation probability as a function of the ratio $L/E_\nu$}. $E_\nu$ was estimated by measuring the muon energy, $E_\mu$, by means of  muon Multiple Coulomb Scattering (MCS) in the rock absorbers in the lower MACRO. The space resolution achieved was \(  \simeq 3{\mm } \). The distribution of the ratio $R = (Data/MC_{no osc})$ obtained
  by this analysis is plotted in Fig. \ref{fig3} versus $(L/E_\nu)$ \cite{r7}. 
\begin{figure}[th]
\begin{center}
\resizebox{0.4\textwidth}{!}{%
  \includegraphics{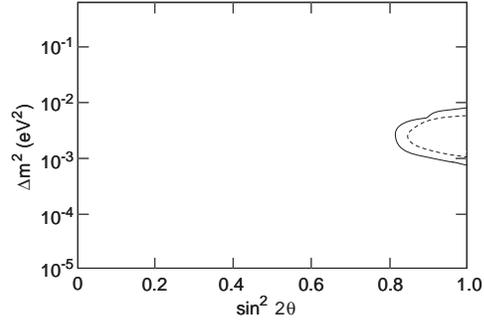} }
\end{center}
\vspace{-1.2cm}
\caption{\label{fig5}\small  
Interpolated qualitative 
90\% C.L. contour plots of the allowed regions  for the MACRO data using only the ratios $R_{1},
R_{2}, R_{3}$ (outer continuous line)  and using also the absolute values
assuming the validity of the Bartol96 fluxes (dotted line). 
}
\end{figure}

The \textit{Internal Upgoing (IU) muons} come from  $\sim 3$ GeV $ \nu _{\mu }$'s interacting in the lower
apparatus. Compared to the no-oscillation prediction there is a reduction in
the flux, without distortion in the   zenith
distribution shape, Fig. \ref{fig2}b. The MC predictions for no oscillations in
Fig. \ref{fig2}b is  the dashed line with a $21
\:\%$ systematic   band. 

The \textit{Upstopping (UGS) muons}, due to $\sim 3$ GeV  \( \nu _{\mu } \)'s
interacting below the detector, yield upgoing muons stopping in the
detector. The \textit{Semicontained Downgoing } (ID) \(\mu\)'s are due to \( \nu
_{\mu } \)-induced downgoing $\mu$'s with vertex in the lower MACRO. 
The 2 types of events (not shown in Fig. \ref{fig2}) are identified by 
topological criteria. 
\section{Oscillation parameters.}
\label{sec:2}
In the past, in order to determine the oscillation parameters, we made fits
to the shape of the upthroughgoing muon zenith distribution and to the absolute
flux compared to the Bartol96 MC. The other data were only used to verify the consistency
and  make checks. The result  was $\Delta m^{2} = 0.0025$ eV$^2$ and
maximal mixing \cite{r6} \cite{r3}. Later, also the $L/E_\nu$ distribution was used \cite{r7}.

In order to reduce the effects of  systematic uncertainties in the MC we recently used the following three independent ratios \cite{r8} and we checked that FLUKA, Honda and Bartol96 MC simulations yield the same predictions to within $\sim 5 \%$.
\begin{enumerate}
        \item [(i)] High Energy Data: zenith distribution ratio: $R_{1} = N_{vert}/N_{hor}$
        \item [(ii)] High Energy Data, neutrino energy measurement ratio: $R_{2} = N_{low}/N_{high}$
        \item [(iii)] Low Energy Data: \\ratio $R_{3} = (Data/MC)_{IU}/(Data/MC)_{ID+UGS}$.
\end{enumerate}
With these ratios, the no oscillation hypothesis has a probability $P \sim 3 \cdot 10^{-7} $
and is  ruled out by  $ \sim 5 \sigma$.
By fitting the 3 ratios  to the \numu $\rightarrow$ \nutau oscillation
formulae we  obtain $\sin^{2} 2\vartheta = 1$,   $\Delta  m^{2} = 2.3
\cdot 10^{-3}$ eV$^{2}$ and the allowed region indicated by the solid line in Fig. \ref{fig5}.

If we use the Bartol96 flux we may add   the information on the absolute fluxes of the
\begin{enumerate}
   \item[(iv)] high energy data (systematic scale error of $\gsim 17 \%$) $R_{4} =
     N_{meas}/N_{MCBartol}$.
   \item [(v)] low energy semicontained muons, with a systematic scale
     error of $21 \%$, $R_{5} \simeq N_{meas}/N_{MCBartol}$.
     \end{enumerate}
These informations reduce  the area of the allowed region  (dashed line
in Fig. \ref{fig5}), do not change the best fit values and bring the significance to $\sim 6 \sigma$.

\begin{figure}[th]
\begin{center}
\resizebox{0.9\hsize}{!}{
  \includegraphics{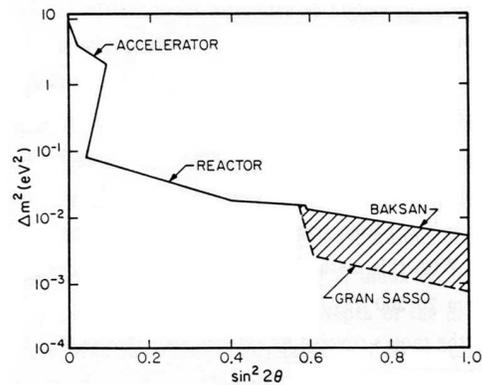}
  }
\end{center}
\vspace{-1.2cm}
\caption{\label{figmacro}\small From the 1984 MACRO proposal: 1984 limits on $\Delta m^2,\; sin^2 2\vartheta$. The shaded region indicated the expected improvement in the explored region obtainable with the MACRO experiment. }
\end{figure}

We recall that in the 1984  proposal it was stressed that the new region in $(\Delta m^{2},\;
\sin^{2}2\theta)$ that MACRO would cover is as indicated in Fig. \ref{figmacro}, where the shape of the angular distribution of upthroughgoing \(\mu\)'s is very sensitive.
\begin{figure}[th]
\begin{center}
\resizebox{0.9\hsize}{!}{%
  \includegraphics{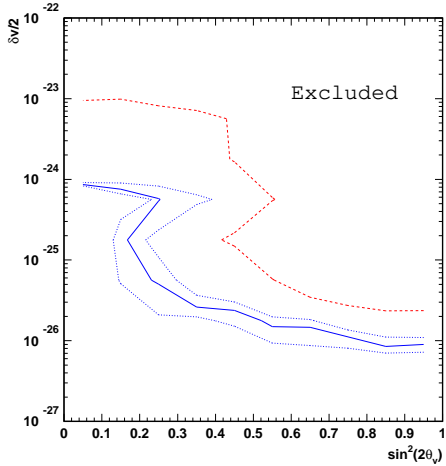}
  }
\end{center}
\vspace{-1.2cm}
\caption{\label{liv}\small 90\%C.L. upper limits on the LIV parameters $\Delta v$ and sin$^2 2\vartheta_v$, assuming $\Delta m^2 =0.0023 \;eV^2$, sin$^2 2\vartheta_m$ =1 for the mass induced oscillations (solid line). The upper (lower) dotted line refers to $\Delta m^2$ = 0.0015 (0.0034) eV$^2$. The dashed line shows the limit obtained using the selection criteria of ref. \cite{r7} }
\end{figure}
\section{Search for possible Lorentz invariance violation contributions.}
Two flavor \( \nu _{\mu }\rightarrow  \nu _{\tau } \) mass-induced oscillations are strongly favored over a wide range of alternative explanations of the atmospheric $\nu$ data \cite{r14} \cite{r15}. In a search for possible Lorentz invariance violation (LIV) contributions, we assumed mass-induced $\nu$ oscillations as the leading mechanism and LIV as a sub-dominant effect \cite{r16}. In this scenario one  considers 2 flavor eigenstates, 2 mass eigenstates and 2 velocity eigenstates (characterized by different maximum attainable velocities in the limit of infinite momentum). In a first analysis we considered the subsample of upthroughgoing $\mu$'s for which the energy was estimated via MCS and we used the ratio  $R'_2 = N'_{low}/N'_{high}$, where low and high are for  events with reconstructed energies $E_\nu^{rec}<$  30 and  $>$ 130  GeV (average energies of 13  and 146 GeV, respectively). In the analysis we fixed the neutrino mass oscillation parameters at $\Delta m^2 = 0.0023\; eV^2$ and $\sin^{2} 2\vartheta_m = 1$ and performed a minimization with respect to the LIV parameters $\Delta v$ and $\vartheta_v$. The inclusion of LIV effects does not improve the $\chi^2$ of the fit and we obtain the $90 \%$ C.L. limits shown as the continuous line in Fig. \ref{liv}. The calculation was repeated for median neutrino energies of 17 and 167 GeV,  dashed line in Fig. \ref{liv}.\\

We would like to acknowledge the cooperation of  the members of the MACRO collaboration and discussions with several theoretical colleagues.


\begin{thebibliography}{}
\bibitem{r1} S. Ahlen et al., 
Nucl. Instr. Meth.  \textbf{A324}(1993)337.
M. Ambrosio et al., 
Nucl. Instr. Meth. \textbf{A486}(2002)663. 
\bibitem{r2} G. Giacomelli et al., 
hep-ex/0211035;
hep-ex/0210006; 
Modern Phys. Lett. \textbf{A18}(2003)2001; hep-ex/0407023.
\bibitem{r3} M. Ambrosio et al., 
Astrop. Phys. \textbf{9}(1998)105;  hep-ex/0206027.
\bibitem{r4} S. Ahlen et al., 
Phys. Lett.  \textbf{B357}(1995)481.
M. Ambrosio et al., 
Phys. Lett. \textbf{B434} (1998) 451; 
Phys. Lett. \textbf{B478}(2000)5.
\bibitem{r6} M. Ambrosio et al., 
Phys. Lett. \textbf{B517}(2001)59.
\bibitem{r7} M. Ambrosio et al., 
Nucl. Instr. Meth.  \textbf{A492}(2002)376;
  Phys. Lett. \textbf{B566}(2003)35.
\bibitem{r8} M. Ambrosio et al., 
Eur. Phys. J. C36(2004)323.
\bibitem{bartol} V. Agrawal et al., Phys. Rev. \textbf{D53}(1996)1314.  
\bibitem{honda95} M. Honda et al., Phys. Rev. \textbf{D52} (1995) 4985. 
\bibitem{sk98} Y. Fukuda et al., Phys. Rev. Lett. 81 (1998) 1562.       
\bibitem{honda01} M. Honda et al., Phys. Rev. \textbf{D64}(2001) 053011; Phys. Rev. D70 (2004) 043008. 
\bibitem{fluka} G. Battistoni et al., Astrop. Phys. \textbf{19}(2003)269; Astrop. Phys. \textbf{19}(2003)291.
\bibitem{hayato} Y. Ashie et al., Phys. Rev. Lett. 93 (2004) 101801. 
\bibitem{r13} L3C Coll., P. Le Coultre et al.; BESS Coll., T. Sanaki et al., Talks at this workshop.
\bibitem{r14} S. Coleman and S. L. Glashow, Phys. Lett. B405(1997) 249; Phys. Rev. D59  (1999) 116008.
S. L. Glashow, hep-ph/0407087.
\bibitem{r15} G. L. Fogli et al. Phys. Rev. D60 (1999) 053006.
\bibitem{r16} G. Battistoni et al., hep-ex/0503015.
\end{thebibliography}
\end{document}